\documentclass{desyproc}
\begin{document}
\title{First Axion Dark Matter Search with Toroidal Geometry}

\author{{\slshape  Byeong Rok Ko}\\[1ex]
  Center for Axion and Precision Physics Research (CAPP), Institute for Basic
  Science (IBS), \\
  Daejeon 34141, Republic of Korea}

\contribID{familyname\_firstname}

\confID{13889}  
\desyproc{DESY-PROC-2017-XX}
\acronym{Patras 2017} 
\doi  

\maketitle

\begin{abstract}
  \hspace{3ex}
  We report the first axion dark matter search with toroidal
  geometry. Exclusion limits of the axion-photon coupling
  $g_{a\gamma\gamma}$ over the axion mass range from 24.7 to 29.1
  $\mu$eV at the 95\% confidence level are set through this pioneering
  search. Prospects for axion dark matter searches with larger scale
  toroidal geometry are also given.
\end{abstract}

\section{Introduction}
\begin{wrapfigure}{r}{0.45\textwidth}
  \centerline{\includegraphics[width=0.45\textwidth]{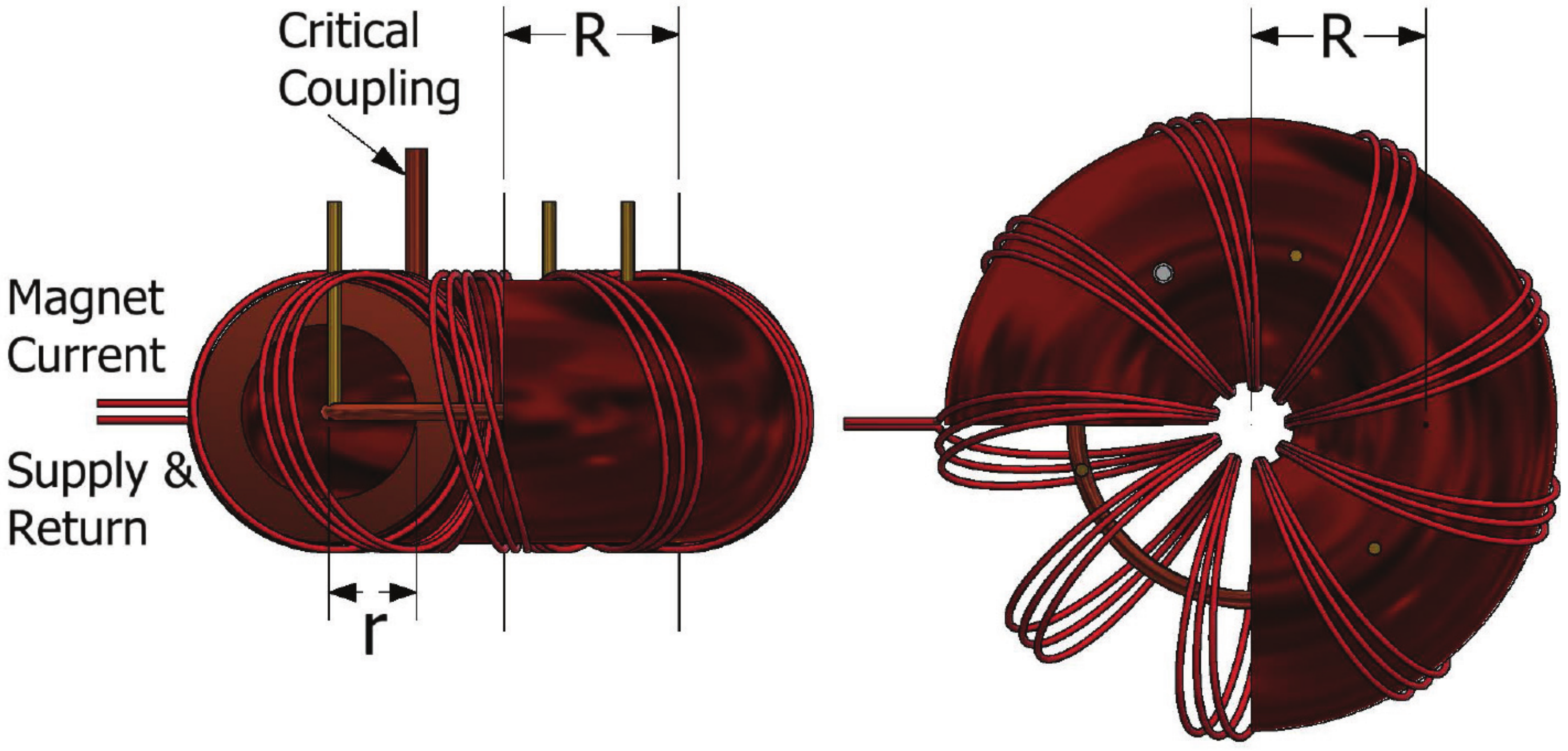}}
  \caption{Lateral (left) and top (right) views of the CAPPuccino
    submarine. Note that it is a cut-away view to show details of the
    system.}
  \label{FIG:CAPPUCCINO}
\end{wrapfigure}

\hspace{3ex}
In the last Patras workshop at Jeju Island in Republic of Korea, we,
IBS/CAPP, introduced axion haloscopes with toroidal geometry we will
pursue~\cite{BRKo_PATRAS2016}. At the end of our presentation, we
promised that we will show up at this Patras workshop with
``CAPPuccino submarine''. The CAPPuccino submarine is a copper
(cappuccino color) toroidal cavity system whose lateral shape is
similar to a submarine as shown in Fig.~\ref{FIG:CAPPUCCINO}.

\noindent We are now referring to the axion dark matter searches with
toroidal geometry at our center as ACTION for ``{\bf A}xion haloscopes
at {\bf C}APP with {\bf T}oro{\bf I}dal res{\bf ON}ators'' and the
ACTION in this proceedings is the ``simplified ACTION''. In this
proceedings, we mainly show the first axion haloscope search from the
simplified ACTION experiment and also discuss the prospects for larger
scale ACTION experiments~\cite{SIMPLE_ACTION}.

\section{Simplified ACTION}
\hspace{3ex}
The simplified ACTION experiment constitutes a tunable copper toroidal
cavity, toroidal coils which provide a static magnetic field, and a
typical heterodyne receiver chain. The experiment was conducted at
room temperature. A torus is defined by $x=(R+r\cos\theta)\cos\phi$,
$y=(R+r\cos\theta)\sin\phi$, and $z=r\sin\theta$, where $\phi$ and
$\theta$ are angles that make a full circle of radius $R$ and $r$,
respectively. As shown in Fig.~\ref{FIG:CAPPUCCINO}, $R$ is the
distance from the center of the torus to the center of the tube and
$r$ is the radius of the tube. Our cavity tube's $R$ and $r$ are 4 and
2 cm, respectively, and the cavity thickness is 1 cm.
\begin{wrapfigure}{r}{0.45\textwidth}
  \centerline{\includegraphics[width=0.5\textwidth]{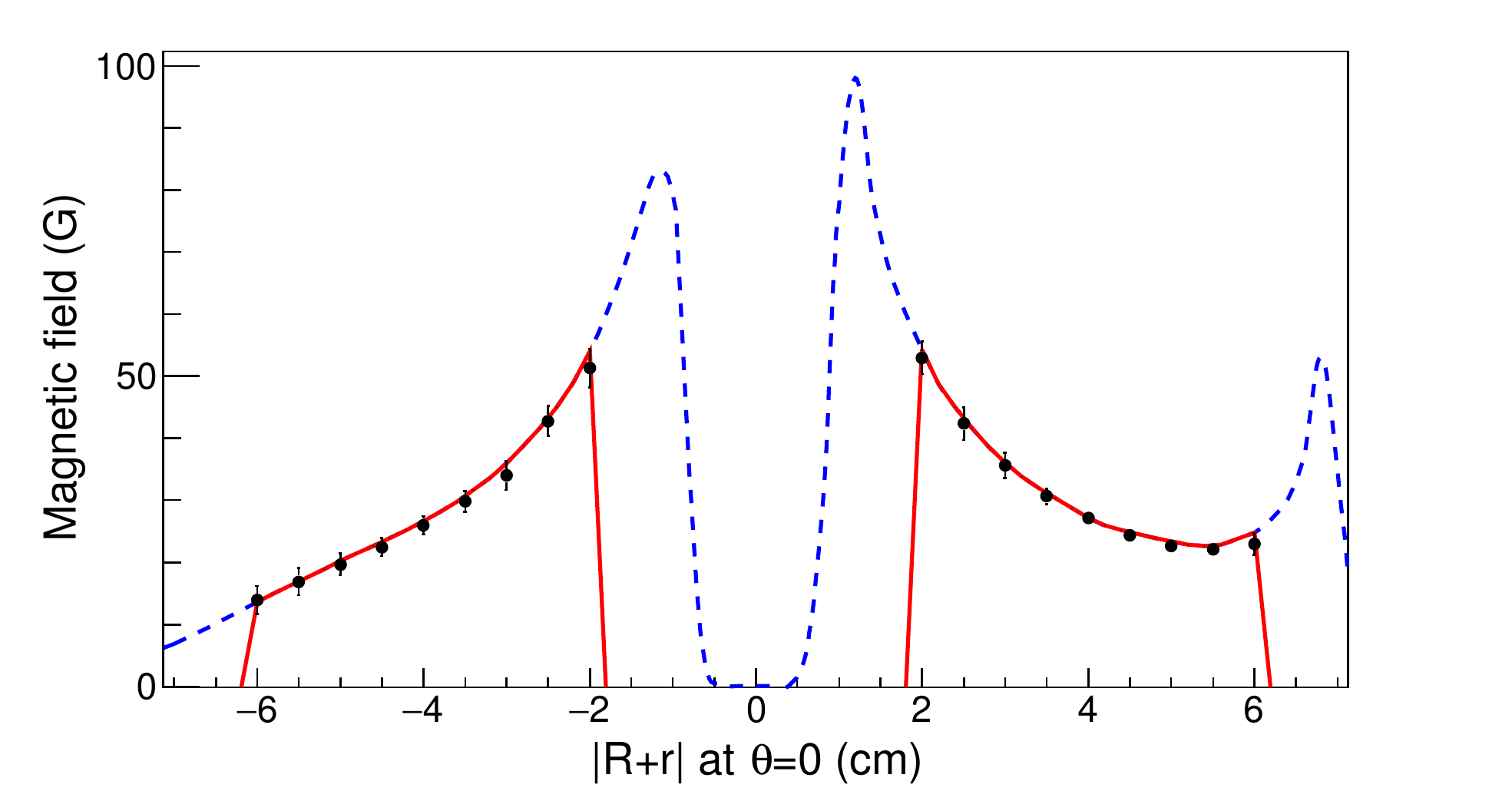}}
  \caption{Magnetic field as a function of radial position $|R+r|$ at
    $\theta=0$. Dashed (blue) lines are obtained from the finite
    element method and correspond to the toroidal cavity system, and
    solid lines (red) correspond to the cavity tube. Dots with error
    bars are measurement values. The results at positive $R+r$ are
    along a coil, while those at negative $R+r$ are between two
    neighboring coils.}  
  \label{FIG:BSIMULATION}
\end{wrapfigure}

\noindent The frequency tuning system constitutes a copper tuning hoop
whose $R$ and $r$ are 4 and 0.2 cm, respectively, and three brass
posts for linking between the hoop and a piezo linear actuator that
controls the movement of our frequency tuning system. The
quasi-TM$_{010}$ (QTM$_1$) modes of the cavity are tuned by moving up
and down our frequency tuning system along the axis parallel to the
brass posts. Two magnetic loop couplings were employed, one for weakly
coupled magnetic loop coupling and the other for critically coupled
magnetic loop coupling, i.e. $\beta\simeq1$ to maximize the axion
signal power in axion haloscope searches~\cite{sikivie}.
\begin{wrapfigure}{r}{0.45\textwidth}
  \centerline{\includegraphics[width=0.5\textwidth]{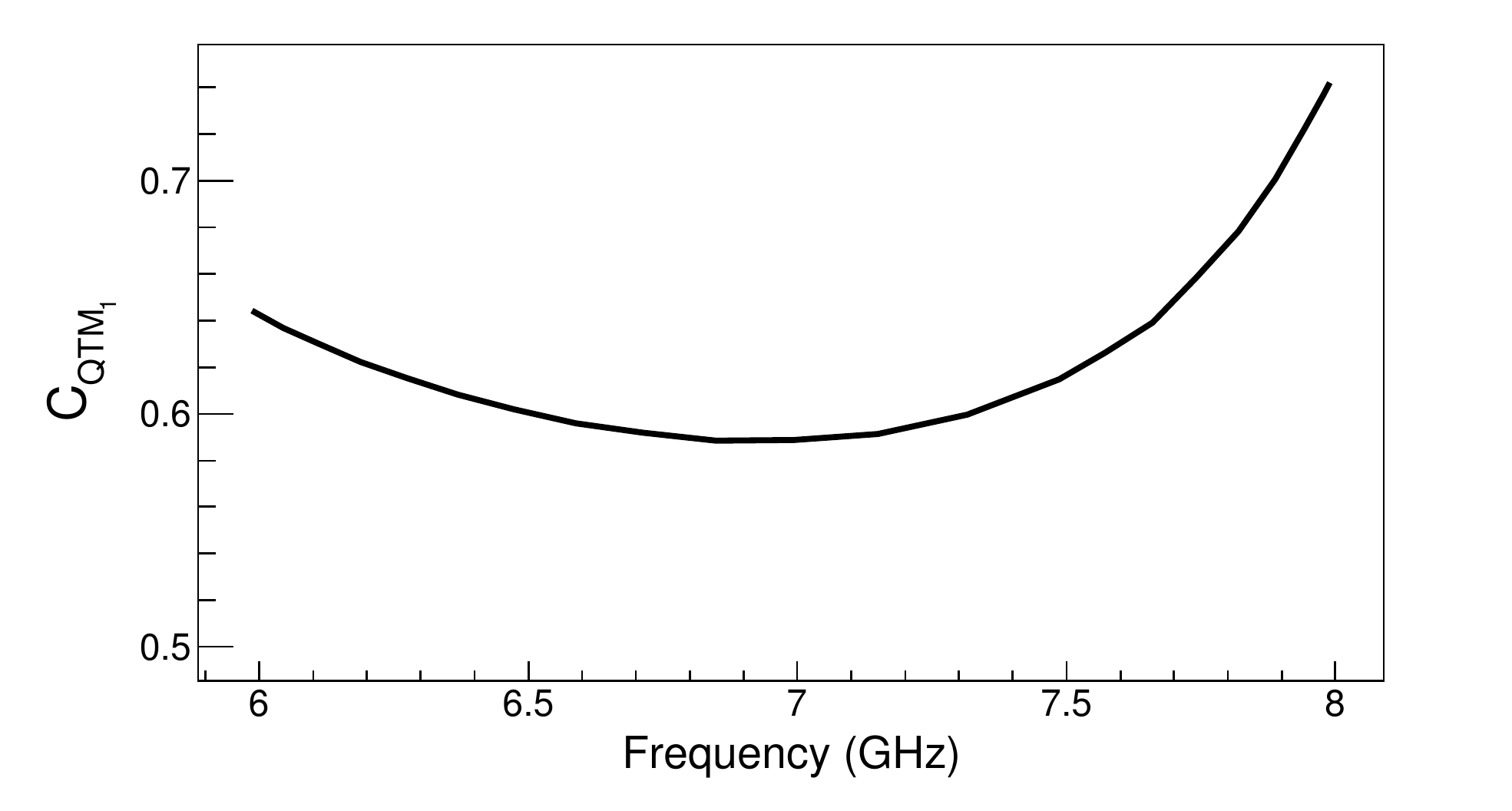}}
  \caption{Form factors of the QTM$_1$ mode of the toroidal cavity as
    a function of the QTM$_1$ frequency.}
  \label{FIG:FF}
\end{wrapfigure}

A static magnetic field was provided by a 1.6 mm diameter copper wire
ramped up to 20 A with three winding turns, as shown in
Fig.~\ref{FIG:CAPPUCCINO}. Figure~\ref{FIG:BSIMULATION} shows good
agreement between measurement with a Hall probe and a
simulation~\cite{CST} of the magnetic field induced by the coils. The
$B_{\rm avg}$ from the magnetic field map provided by the simulation
turns out to be 32 G.

With the magnetic field map and the electric field map of the QTM$_1$
mode in the toroidal cavity, we numerically evaluated the form factor
of the QTM$_1$ mode as a function of the QTM$_1$ frequency, as shown in
Fig.~\ref{FIG:FF}, where the highest frequency appears when the
frequency tuning system is located at the center of the cavity tube,
such as in Fig.~\ref{FIG:CAPPUCCINO}. As shown in Fig.~\ref{FIG:FF},
we found no significant drop-off in the form factors of the QTM$_1$
modes, which is attributed to the absence of the cavity endcaps in
toroidal geometry.

Our receiver chain consists of a single data acquisition channel that
is analogous to that adopted in ADMX~\cite{ADMX_NIM} except for the
cryogenic parts. Power from the cavity goes through a directional
coupler, an isolator, an amplifier, a band-pass filter, and a mixer,
and is then measured by a spectrum analyzer at the end. Cavity
associates, $\nu$ (resonant frequency), and $Q_L$ (quality factor with
$\beta\simeq1$) are measured with a network analyzer by toggling
microwave switches. The gain and noise temperature of the chain were
measured to be about 35 dB and 400 K, respectively, taking into
account all the attenuation in the chain, for the frequency range from
6 to 7 GHz.

The signal-to-noise ratio (SNR) in the simplified ACTION experiment is
\begin{equation}
  {\rm SNR}=\frac{P_{a,g_{a\gamma\gamma}\sim6.5\times10^{-8}~{\rm GeV}^{-1}}}{P_n}\sqrt{N},
  \label{EQ:SNR}
\end{equation}
where $P_{a,g_{a\gamma\gamma}\sim6.5\times10^{-8}~{\rm GeV}^{-1}}$ is
the signal power when $g_{a\gamma\gamma}\sim6.5\times10^{-8}$
GeV$^{-1}$, which is approximately the limit achieved by the ALPS
collaboration~\cite{ALPS2010}. $P_n$ is the noise power equating to
$k_B T_n b_a$, and $N$ is the number of power spectra, where $k_B$ is
the Boltzmann constant, $T_n$ is the system noise temperature which is
a sum of the noise temperature from the cavity ($T_{n,{\rm cavity}}$)
and the receiver chain ($T_{n,{\rm chain}}$), and $b_a$ is the signal
bandwidth. We iterated data taking as long as $\beta\simeq 1$, or
equivalently, a critical coupling was made, which resulted in about 3,500
measurements. In every measurement, we collected 3,100 power spectra and
averaged them to reach at least an SNR in Eq.~(\ref{EQ:SNR}) of about
8, which resulted in an SNR of 10 or higher after overlapping the
power spectra at the end.
\begin{wrapfigure}{r}{0.45\textwidth}
  \centerline{\includegraphics[width=0.45\textwidth]{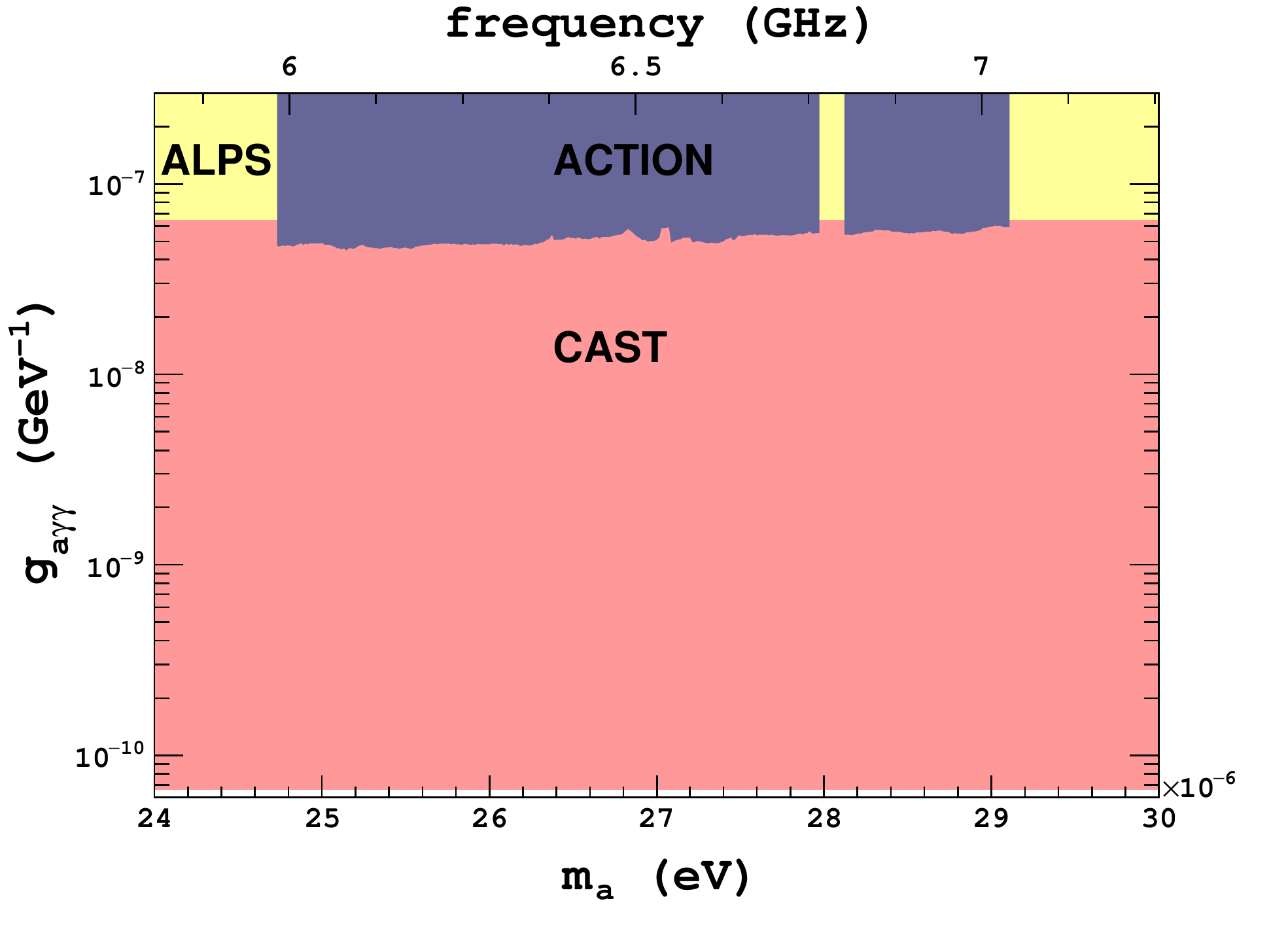}}
  \caption{Excluded parameter space at the 95\% C.L. by this
    experiment together with previous results from
    ALPS~\cite{ALPS2010} and CAST~\cite{CAST2017}. No limits are set
    from 6.77 to 6.80 GHz due to with a quasi-TE mode in that
    frequency region and the TE mode is also confirmed by a
    simulation~\cite{CST}.}
  \label{FIG:LIMIT}
\end{wrapfigure}

Our overall analysis basically follows the pioneer study described in
Ref.~\cite{haloscope4}. With an intermediate frequency of 38 MHz,
we take power spectra over a bandwidth of 3 MHz, which allows 10 power
spectra to overlap in most of the cavity resonant frequencies with a
discrete frequency step of 300 kHz. Power spectra are divided by the
noise power estimated from the measured and calibrated system noise
temperatures. The five-parameter fit also developed in
Ref.~\cite{haloscope4} is then employed to eliminate the residual
structure of the power spectrum.
The background-subtracted power spectra are combined in order to
further reduce the power fluctuation. We found no significant excess
from the combined power spectrum and thus set exclusion limits of
$g_{a\gamma\gamma}$ for $24.7<m_a<29.1$ $\mu$eV. No frequency bins in
the combined power spectrum exceeded a threshold of 5.5$\sigma_{P_n}$,
where $\sigma_{P_n}$ is the rms of the noise power $P_n$.
We found $\sigma_{P_n}$ was underestimated due to the five-parameter
fit as reported in Ref.~\cite{HF} and thus corrected for it
accordingly before applying the threshold of 5.5$\sigma_{P_n}$. Our
SNR in each frequency bin in the combined power spectrum was also
combined with weighting according to the Lorentzian lineshape,
depending on the $Q_L$ at each resonant frequency of the cavity.
With the tail of the assumed Maxwellian axion signal shape, the best
SNR is achieved by taking about 80\% of the signal and associate noise
power; however, doing so inevitably degrades SNR in Eq.~(\ref{EQ:SNR})
by about 20\%. Because the axion mass is unknown, we are also unable
to locate the axion signal in the right frequency bin, or
equivalently, the axion signal can be split into two adjacent
frequency bins.
On average, the signal power reduction due to the frequency
binning is about 20\%. The five-parameter fit also degrades the signal
power by about 20\%, as reported in Refs.~\cite{haloscope4, HF}. Taking
into account the signal power reductions described above, our SNR
for $g_{a\gamma\gamma}\sim6.5\times10^{-8}$ GeV$^{-1}$ is greater or
equal to 10, as mentioned earlier. The 95\% upper limits of the power
excess in the combined power spectrum are calculated in units of
$\sigma_{P_n}$; then, the 95\% exclusion limits of $g_{a\gamma\gamma}$
are extracted using $g_{a\gamma\gamma}\sim6.5\times10^{-8}$ GeV$^{-1}$
and the associated SNRs we achieved in this work. Figure~\ref{FIG:LIMIT}
shows the excluded parameter space at a 95\% confidence level (C.L.)
by the simplified ACTION experiment.
\begin{wrapfigure}{r}{0.5\textwidth}
  \centerline{\includegraphics[width=0.5\textwidth]{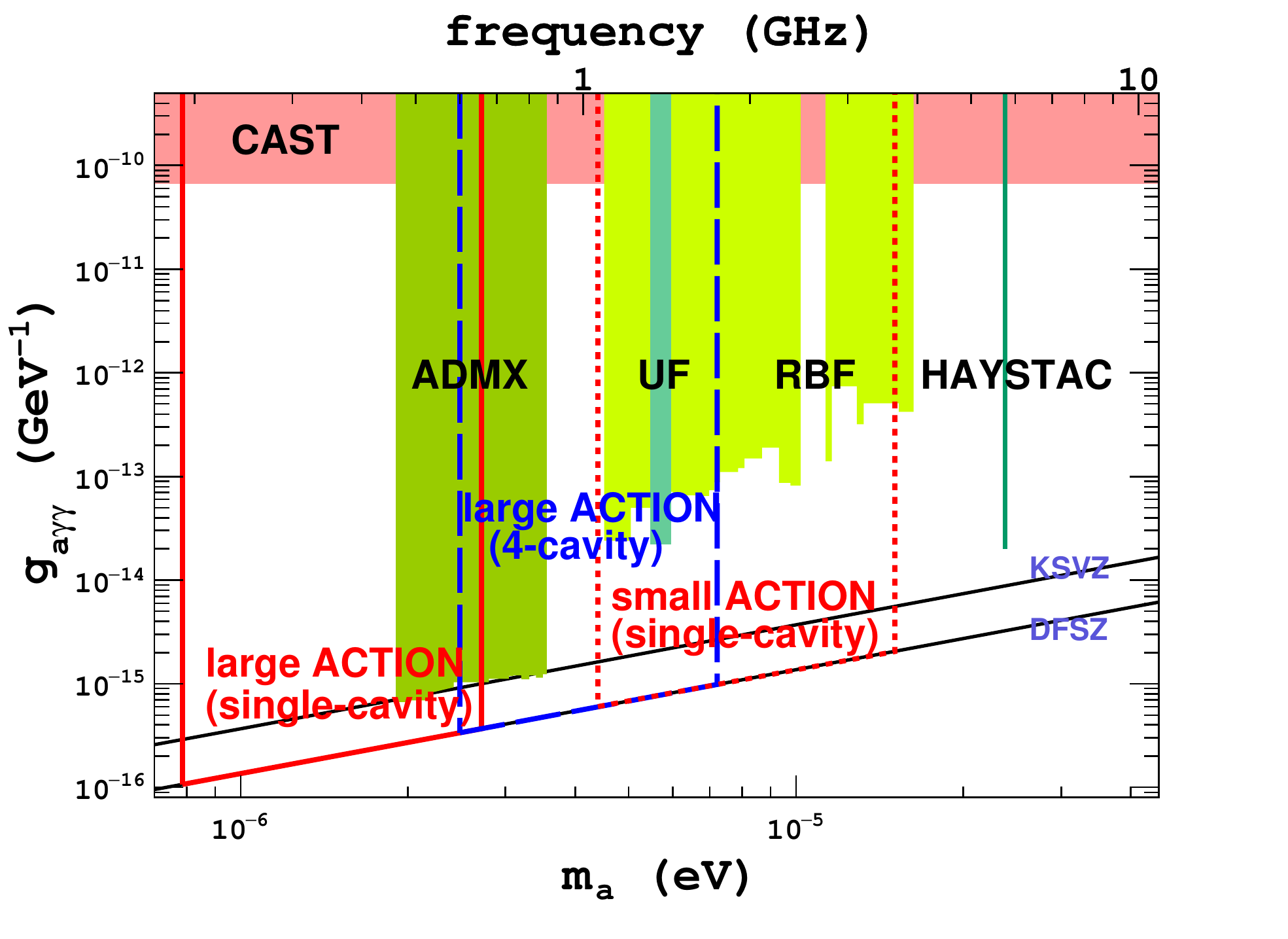}}
  \caption{Expected exclusion limits by the large (solid lines with a
    single-cavity and dashed lines with a 4-cavity) and small (dotted
    lines with a single-cavity) ACTION experiments. Present exclusion
    limits~\cite{haloscope4, HF, CAST2017, haloscope2, haloscope3,
      haloscope5} are also shown.}
  \label{FIG:PROSPECTS}
\end{wrapfigure}
\section{Prospects for axion dark matter searches with larger scale toroidal geometry}
\hspace{3ex}
The prospects for axion dark matter searches with two larger-scale
toroidal geometries that could be sensitive to the
KSVZ~\cite{KSVZ1,KSVZ2} and DFSZ~\cite{DFSZ1,DFSZ2} models are now
discussed. A similar discussion can be found
elsewhere~\cite{DIPOLE}. One is called the ``large ACTION'', and the
other is the ``small ACTION'', where the cavity volume of the former
is about 9,870 L and that of the latter is about 80 L. The $B_{\rm
  avg}$ targets for the large and small ACTION experiments are 5 and
12 T, respectively, where the peak fields of the former and latter
would be about 9 and 17 T. Hence, the large and small toroidal magnets
can be realized by employing NbTi and Nb$_3$Sn superconducting wires,
respectively. The details of the expected experimental parameters for
the ACTION experiments can be found in~\cite{SIMPLE_ACTION} and
Fig.~\ref{FIG:PROSPECTS} shows the exclusion limits expected from the
large and small ACTION experiments.
\section{Summary}
\hspace{3ex}
In summary, we, IBS/CAPP, have reported an axion haloscope search
employing toroidal geometry using the simplified ACTION
experiment. The simplified ACTION experiment excludes the axion-photon
coupling $g_{a\gamma\gamma}$ down to about $5\times10^{-8}$ GeV$^{-1}$
over the axion mass range from 24.7 to 29.1 $\mu$eV at the 95\%
C.L. This is the first axion haloscope search utilizing toroidal
geometry since the advent of the axion haloscope search by
Sikivie~\cite{sikivie}. We have also discussed the prospects for axion
dark matter searches with larger scale toroidal geometry that could be
sensitive to cosmologically relevant couplings over the axion mass
range from 0.79 to 15.05 $\mu$eV with several configurations of tuning
hoops, search modes, and multiple-cavity system.
\section*{Acknowledgments}
\hspace{3ex}
This work was supported by IBS-R017-D1-2017-a00. 
\begin{footnotesize}

\end{footnotesize}


\end{document}